\documentclass[a4paper,11pt,linenumbers]{article}

\usepackage{pos}
\usepackage{comment}
\usepackage{lipsum} %package to generate placeholder text in the following
\usepackage{lineno}
\usepackage{xcolor}

\title{Calculation of the exposure of GRANDProto300 to cosmic rays}

\ShortTitle{GP300 exposure}

\author*[a]{Sei Kato}
\author[a]{Clément Prévotat}
\author[a, b]{Rafael Alves Batista}
\onbehalf{for the GRAND Collaboration{\normalsize \\ \normalfont(a complete list of authors can be found at the end of the proceedings)}\\}

\affiliation[a]{Institut d’Astrophysique de Paris, CNRS UMR 7095, Sorbonne Université, Paris, France}
\affiliation[b]{Laboratoire de Physique Nucléaire et des Hautes Energies, Sorbonne Université, Paris, France}

\emailAdd{sei.kato@iap.fr}
\abstract{
GRANDProto300 is one of the prototype experiments of the Giant Radio Array for Neutrino Detection. It will feature about 300 radio antenna detectors in Xiaodushang in Dunhuang, China, covering a total geometrical area of about $200\, {\rm km}^2$. A main scientific goal of GRANDProto300 is the study of cosmic rays in the transition region ($10^{17}\, {\rm eV} < E < 10^{18.5}\, {\rm eV}$). Our study calculates the exposure of GRANDProto300 to cosmic rays and estimates the number of cosmic-ray events to be detected during a fixed observation period. The trigger efficiency reaches 50, 80, and 90\% at $10^{17.5}$, $10^{17.9}$, and $10^{18.3}\, {\rm eV}$, respectively. The exposure of GRANDProto300 is $50\, {\rm km}^2\, {\rm day}\, {\rm sr}$ at around $10^{17.5}\, {\rm eV}$, and the expected number of observed cosmic rays with energies above $10^{17}\, {\rm eV}$ and zenith angles above $65^{\circ}$ is about $130$ events per day. GRANDProto300 will be able to measure the cosmic-ray energy spectrum in $10^{17.2}\, {\rm eV}<E<10^{19.5}\, {\rm eV}$ through one-year observation, with a statistical precision about five times better than the previous spectral measurement by a mono-fluorescence detector of the Telescope Array Low-Energy Extension. The statistical uncertainty in the measurement of the mean depth of the air-shower maximum $X_{\rm max}$ is about five times better than the previous measurements using radio detectors at $10^{17.5}\, {\rm eV}$; systematic uncertainties should be a dominant contribution limiting our interpretation of the chemical composition of cosmic rays in the transition region.
\vspace{4mm}
}

\ConferenceLogo{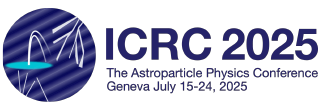}

\FullConference{39th International Cosmic Ray Conference (ICRC2025)\\  15--24 July, 2025\\ Geneva, Switzerland}

\begin{document}

\maketitle
%\linenumbers
%\section{Instructions for Authors}\label{sec1}
\section{Introduction} \label{sec1}
The Giant Radio Array for Neutrino Detection (GRAND) is a future radio-antenna-array experiment mainly aiming at the first detection of ultra-high-energy neutrinos well beyond $10^{17}\, {\rm eV}$ and the study of the origin of ultra-high-energy cosmic rays ($E > 10^{18}\, {\rm eV}$) \cite{GRAND_wp, ARENA_Kotera}. It will detect the geomagnetic radio emission from Earth-skimming hadronic showers produced by the decay of tau leptons, the products of the charged-current reaction of astrophysical tau neutrinos in the Earth. A main technical challenge of GRAND is an autonomous trigger of radio emission from air showers only using radio detectors. GRAND has three prototype experiments, GRANDProto300 in China, GRAND@Auger in Argentina and GRAND@Nançay in France. In particular, GRANDProto300 (hereafter called GP300) will be dedicated to the calibration of our radio detectors, the validation of the autonomous radio trigger, and the study of cosmic rays (CRs) in the transition region ($10^{17}\, {\rm eV} < E < 10^{18.5}\, {\rm eV}$) \cite{ARENA_Chiche}. This study focuses on the ability of GP300 for the CR observation by calculating the exposure of the array to CRs through a simulation-based study.

\section{GRANDProto300 (GP300)} \label{sec2}
\begin{figure}[t!]
%\begin{figure}[h]
\centering
\includegraphics[width=0.9\linewidth]{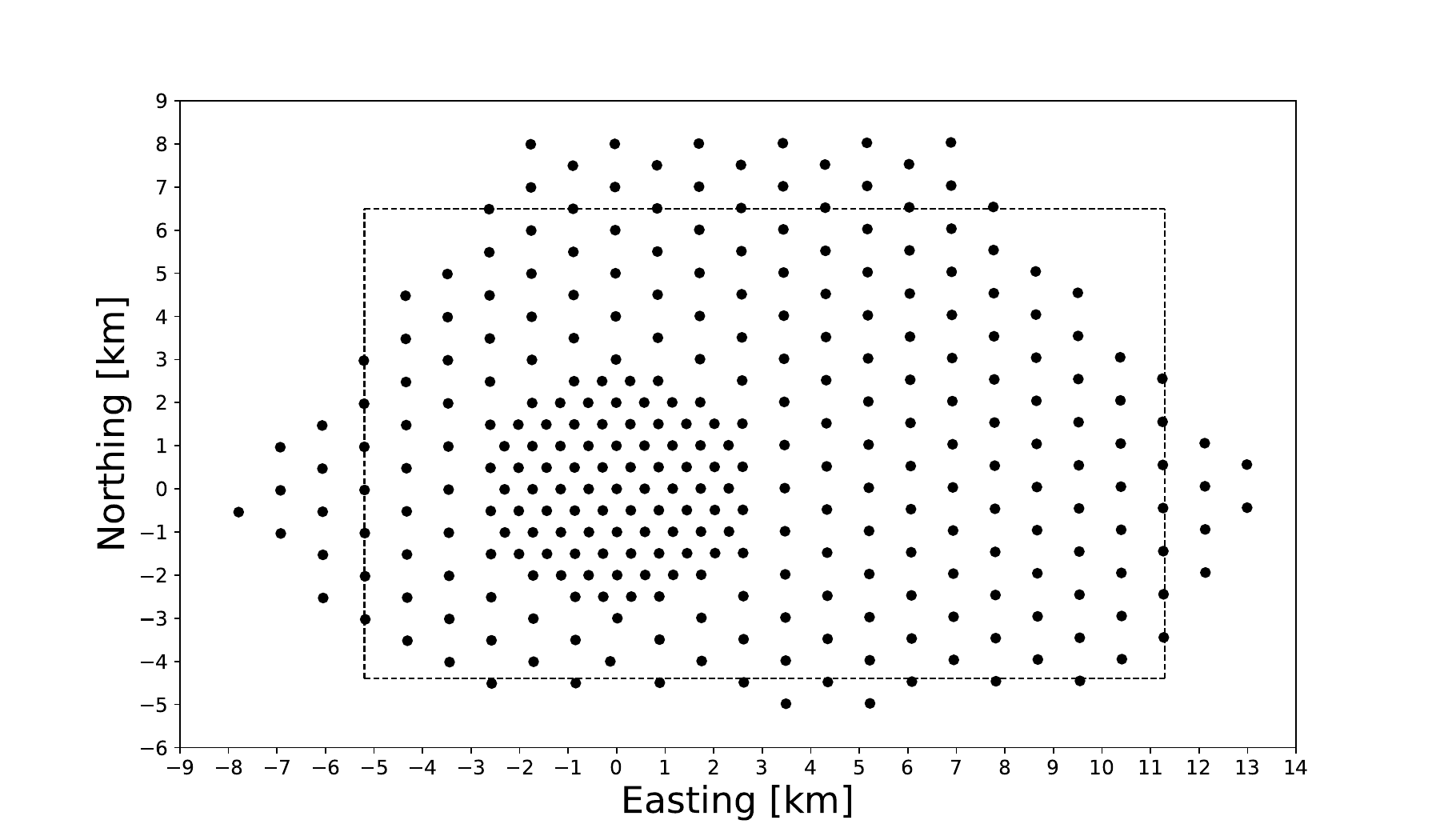}
%\caption{Bird's-eye view of the GP300 radio detector array. Core positions of CR-induced air shower events generated in our simulation are shown with red points. The black dashed lines enclose the rectangular region ([$-$5.2 km, 11.3 km] in Easting and [$-$4.4 km, 6.5 km] in Northing) where the core positions are randomly distributed.}
\caption{Bird's-eye view of the GP300 radio detector array. The black dashed lines enclose the rectangular region ([$-$5.2 km, 11.3 km] in Easting and [$-$4.4 km, 6.5 km] in Northing) where the core positions are randomly distributed.}
\label{GP300-config}
\end{figure}

GP300 is a prototype radio-antenna array to be installed in Xiaodushan ($40.99^{\circ}{\rm N}, 93.94^{\circ}{\rm E}$, $\sim$1200 m above sea level) in Dunhuang, China. Figure \ref{GP300-config} shows the bird's-eye view of GP300. It will make use of 299 radio detector units (DUs) covering a total geometrical area of about 200 ${\rm km}^2$. Each DU has butterfly-shaped steel antenna arms extending in the South-North (X-arm), East-West (Y-arm), and the vertical directions (Z-arm). The X- and Y-arms are dipoles, while the Z-arm is a monopoloe and the supporting pole works as its second arm. For the detailed description of the DU design and its functionality, see \cite{grandlib}. GP300 has an infill array with an interval of $577\, {\rm m}$ between DUs, surrounded by a sparse array with an interval of $1\, {\rm km}$. The infill array is used to lower the energy threshold for CR events. Motivations of GP300 include the validation of the autonomous trigger of radio emission from CR-induced air showers only using radio detectors and the study of the transition region in the CR energy spectrum above $10^{17}\, {\rm eV}$.

\begin{comment}
\begin{figure}[t!]
%\begin{figure}[h]
\centering
\includegraphics[width=0.9\linewidth]{plot_shower_core.pdf}
\caption{Core positions of CR-induced air shower events generated in our simulation (red points), overlaid on the GP300 radio detector array (black points). The black dashed lines enclose the rectangular region ([$-$5.2 km, 11.3 km] in Easting and [$-$4.4 km, 6.5 km] in Northing) where the core positions are randomly distributed.}\label{shower_core}
\end{figure}
\end{comment}

\section{Simulation} \label{sec3}
This study uses the CoREAS code \cite{CoREAS} to simulate the development of CR-induced air showers and calculate the associated radio emission at each position of the GP300 DUs. The CR events are generated following the uniform distribution in the logarithmic energy space within a range of $10^{17}\, {\rm eV} < E < 10^{20}\, {\rm eV}$. The distribution of the incoming directions is uniform in zenith angle $\theta$ within $65^{\circ}<\theta<88^{\circ}$ and azimuthal angle $\phi$ within $0^{\circ}<\phi<360^{\circ}$. Out of total $1.5\times 10^4$ events generated in the simulation, $50\%$ are protons and the remaining $50\%$ are irons. The shower core positions of the events are randomly distributed within the rectangular region enclosed within the black dashed lines shown in Figure \ref{GP300-config}. The region covers only a partial array, so the current analysis using the simulation data only gives an approximate estimate of exposure. The simulation assuming a large geometrical area enough to cover the entire array will be performed in the future.

To calculate the exposure of GP300 for realistic observations, events are weighted to reproduce the measured energy spectrum of CRs and the isotropic distribution of their incoming directions. For the CR energy spectrum, this study assumes the best-fit broken power-law function to the spectral measurement by the fluorescence detector of the Telescope Array Low-Energy Extension (TALE) \cite{Abbasi_2018}:
\begin{equation}\label{tale_sp}
  \frac{{\rm d}N}{{\rm d}E} =
  \begin{cases}
    J_0 \Big(\frac{E}{E_{\rm br}}\Big)^{-2.92} & \text{$10^{17}\, {\rm eV} < E < E_{\rm br}$} \\
    J_0 \Big(\frac{E}{E_{\rm br}}\Big)^{-3.19} & \text{$E_{\rm br} \leq E < 10^{20}\, {\rm eV}$}
  \end{cases}
\end{equation}
where $J_0 = 2.27\times 10^{-16}\, {\rm km}^{-2}\, {\rm day}^{-1}\, {\rm sr}^{-1}\, {\rm eV}^{-1}$ and $E_{\rm br} = 10^{17.04}\, {\rm eV}$. %where $J_0 = 2.6\times 10^{-27}\, {\rm eV}^{-1}\, {\rm m}^{-2}\, {\rm s}^{-1}\, {\rm sr}^{-1}$ and $E_{\rm br} = 10^{17.04}\, {\rm eV}$. %\RAB{[Better write $J_0$ instead of J for the normalisation factor, since J is often used to represent $dN/dE$]}

The radio emission from an air shower event is calculated with CoREAS for the DUs that are located within four Cherenkov radii from the simulated core position of the event. The calculated emission is then processed with \texttt{GRANDlib}~\cite{grandlib} to simulate the output signals from the DUs. The time trace of the electric field of the radio emission is converted into the time trace of open circuit voltage by accounting for the antenna effective length. The open circuit voltage trace is then fed into the radio frequency chain of the DUs to obtain the output voltage trace, which is then digitized by the analog-to-digital (ADC) converter. The Galactic noise: the radio synchrotron emission from high-energy galactic electrons, measured by the real expriment for a fixed observation period \cite{Ma_ICRC2025}, is added to the simulated ADC trace of the shower radio emission. The amplitude of the Galactic noise in our simulation is $15.1$ ADC counts in terms of root mean square (RMS). The value of the noise amplitude is fixed in our simulation analysis, although it fluctuates by a few ADC counts during the operation in the real experiment. For the detailed description of the signal processing in the DUs, see \cite{grandlib}.

To discuss the trigger, the Hilbert envelope of the time trace of ADC counts is calculated in each of the three arms of the DUs. The three traces of the Hilbert envelope are summed in quadrature, and each DU is triggered when the quadrature sum exceeds $90.6$ ADC counts. %which is six times higher than RMS of the added Galactic noise. 
Finally, an overall trigger is issued when any five or more DUs are triggered.

The exposure is calculated from the following equation:
\begin{equation}\label{calc_exposure}
    {\rm Exposure} = A_{\rm geo}\, \Omega\, T_{\rm obs} \frac{\sum_{i}N_{{\rm trig}, i}\,{\rm cos\theta}}{\sum_{j}N_{{\rm sim}, j}},
\end{equation}
where $A_{\rm geo} = 180\, {\rm km}^2$ is the geometrical area of the region where the shower core positions are distributed (see Figure \ref{GP300-config}), $\Omega = 2.4\, {\rm sr}$ the solid angle of the entire sky region assumed in the simulation, and $T_{\rm obs}$ the observation period. $N_{\rm trig}$ is the number of events that triggered the GP300 array, and $N_{\rm sim}$ is the number of events generated in the simulation. Note that the both numbers of events are weighted as mentioned above so that the generated events follow the TALE CR energy spectrum and the isotropic distribution of incoming direction. The ${\rm cos}\, \theta$ factor in the numerator accounts for the fact that the effective geometrical area is smaller by ${\rm cos}\, \theta$ for an injected CR event with a zenith angle of $\theta$.

\section{Results and discussions} \label{sec4}
\begin{figure}[t!]
%\begin{figure}[h]
\centering
\makebox[0pt]{\includegraphics[width=1.2\linewidth]{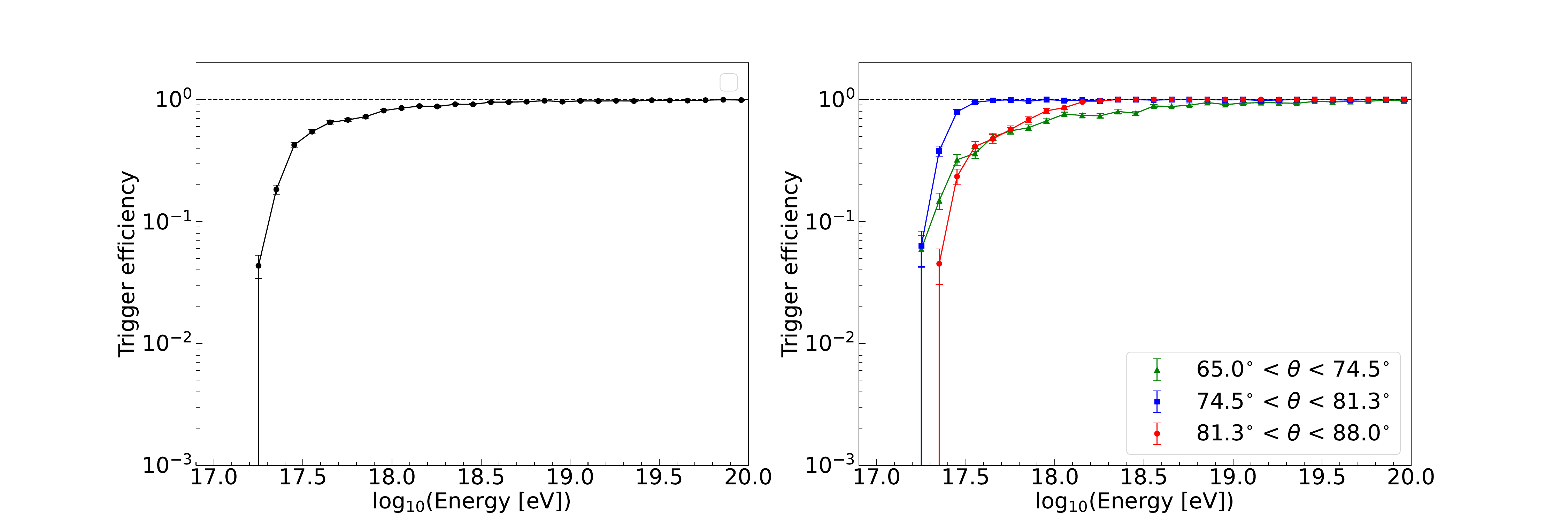}}
\caption{{\it Left panel}: trigger efficiency as a function of energy of CR events. The error bars are statistical errors assuming a binomial distribution for triggered events. The horizontal black dashed line indicates a 100\% efficiency. {\it Right panel}: trigger efficiency for three different zenith-angle ranges. }\label{trig_eff}
\end{figure}

The left panel of Figure \ref{trig_eff} shows the trigger efficiency of GP300 as a function of energy. The trigger efficiency is defined as a fraction of events that triggered the array to the total events generated in the simulation. The error bars show the statistical errors coming from the statistical uncertainty in the number of triggered events assuming a binomial distribution. The horizontal axis shows true energies of CR events generated in the simulation. The trigger efficiency of GP300 reaches 50, 80, and 90\% at $10^{17.5}$, $10^{17.9}$, and $10^{18.3}\, {\rm eV}$, respectively. The right panel of Figure \ref{trig_eff} shows the trigger efficiency for three different zenith angle ranges: $65.0^{\circ} < \theta < 74.5^{\circ}$, $74.5^{\circ} < \theta < 81.3^{\circ}$ and $81.3^{\circ} < \theta < 88.0^{\circ}$. As the zenith angle increases, the trigger efficiency improves due to the large footprint size of the radio emission on the ground, which is able to trigger more DUs. On the other hand, threshold energy becomes higher with an increase of $\theta$ due to the attenuation of the radio emission from highly inclined showers. The figure shows that the efficiency is determined by the balance between the above two physical processes: the footprint size of the air-shower radio emission and its attenuation as a function of zenith angle.

\begin{figure}[h]
\centering
%\includegraphics[width=1.2\linewidth]{calculate_exposure_trig_6sigma.pdf}
%\makebox[0pt]{\includegraphics[width=1.2\linewidth]{calculate_exposure_with_ADCtrace.pdf}}
\makebox[0pt]{\includegraphics[width=1.2\linewidth]{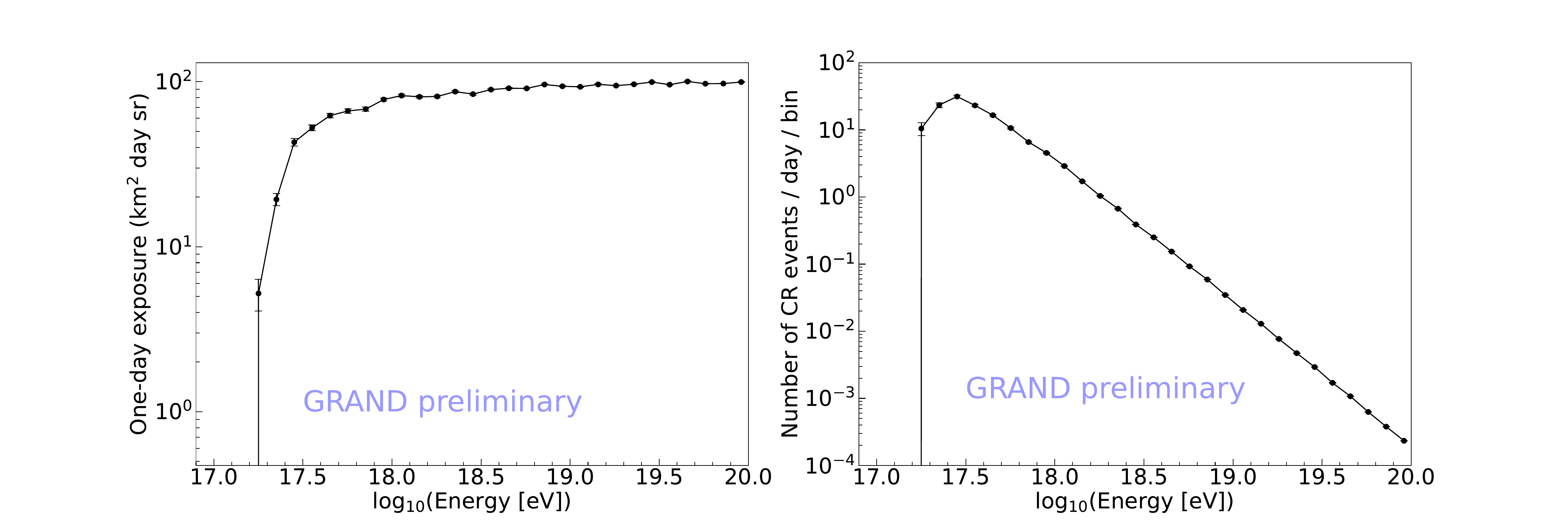}}
\caption{{\it Left panel}: exposure of GP300 in one-day observation as a function of energy of CR events. The error bars are statistical errors assuming a binomial distribution for triggered events. {\it Right panel}: expected number of CR events observed by GP300 per day per energy bin.}\label{exposure}
\end{figure}

The left panel of Figure \ref{exposure} shows the exposure of GP300 as a function of energy for one-day observation calculated from Equation \eqref{calc_exposure} with $T_{\rm obs} = 1\, {\rm day}$. The exposure reaches $50$, $80$, and $90\, {\rm km}^2\, {\rm day}\, {\rm sr}$ at $10^{17.5}$, $10^{18}$, and $10^{19}\, {\rm eV}$, respectively. The right panel of Figure \ref{exposure} shows the expected number of CR events observed by GP300 per day for each energy bin. Integration over the whole energy range in our simulation ($10^{17}\, {\rm eV} < E < 10^{20}\, {\rm eV}$ and $65^{\circ} < \theta < 88^{\circ}$) gives about $130$ events per day. 

%\begin{figure}[t!]

\begin{figure}[t!]
%\begin{figure}[h]
\centering
\includegraphics[width=1\linewidth]{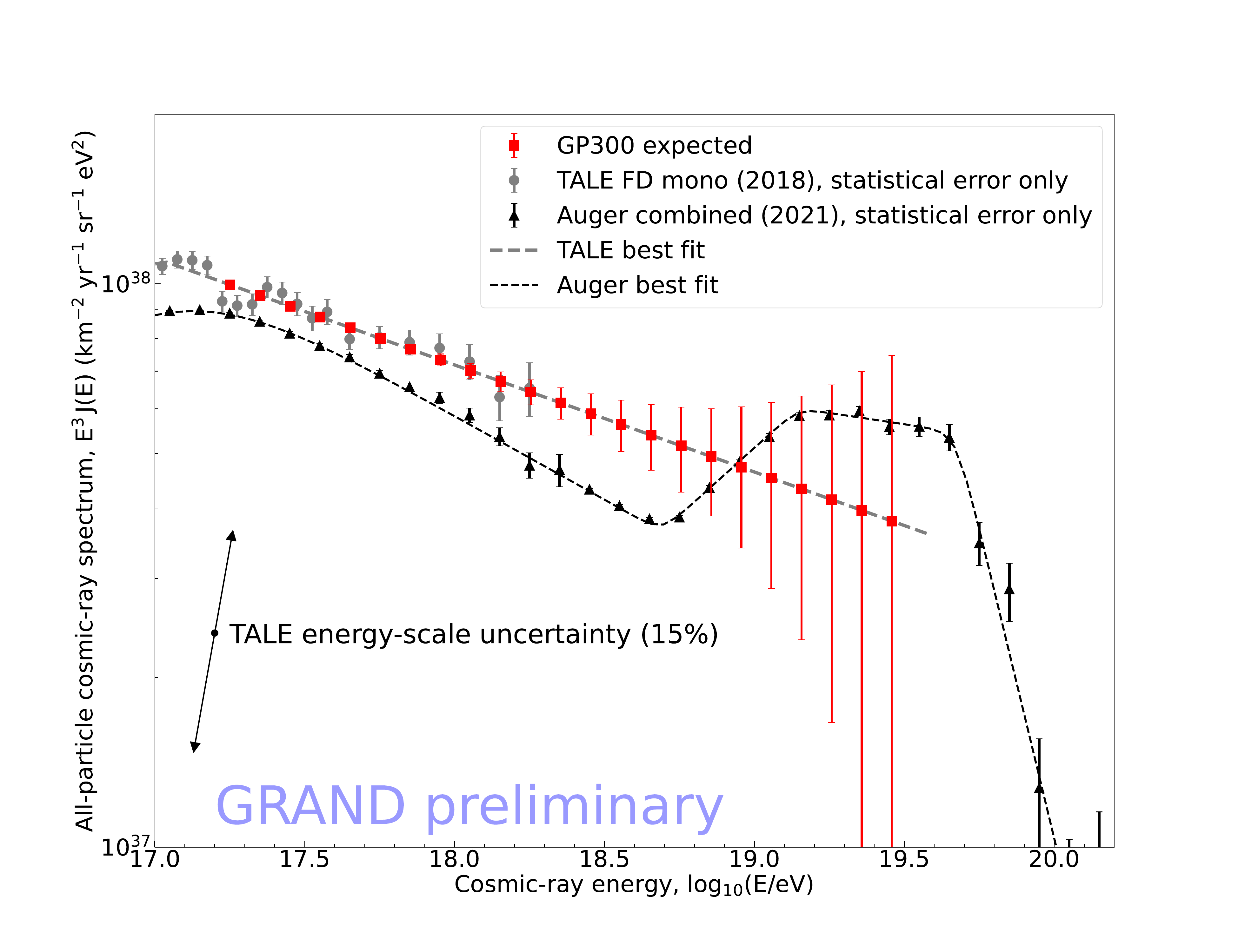}
\caption{Prediction of the measurement of the CR energy spectrum by GP300 for a one-year observation. The error bars show statistical errors assuming a Poisson distribution. The gray data points show the measurement by the TALE fluorescence detector for 1,080 hours of observation \cite{Abbasi_2018}. The gray dashed lines show the best-fit broken power-law function to the TALE measurement; see Equation \eqref{tale_sp}. The black data points are given by the Pierre Auger Observatory, combining the measurements by their surface detector arrays with 750 and 1500 m intervals \cite{Abreu2021}. The error bars of the TALE and Auger measurements are statistical only, and the systematic uncertainty in the flux normalization from the energy-scale uncertainty is indicated with the arrows for the case of TALE.}\label{CRspectrum_prediction} 
%\RAB{[Can't we overlay also markers for Auger, even if we don't use them in the analysis?]}}\label{CRspectrum_prediction}
\end{figure}

Figure \ref{CRspectrum_prediction} shows a prediction of the reconstructed CR spectrum from the one-year observation by GP300 compared with the measurement by the TALE fluorescence detector \cite{Abbasi_2018}. For our prediction, the horizontal axis presents true energies of CR events generated in the simulation. A total number of $4.9\times 10^{4}$ CRs will be observed, and the preliminary estimate of the number of CRs in each energy bin is presented in Table \ref{tab1}. Compared with the previous TALE observation comprehensively covering the transition region, GP300 will be able to determine the CR flux with a better statistical precision by a factor of about eight and five at the $10^{17.5}\, {\rm eV}$ and $10^{18}\, {\rm eV}$ energy ranges, respectively. In the energy range of $10^{18}\, {\rm eV} \lesssim E \lesssim 10^{19}\, {\rm eV}$, the event statistics estimated for the GP300 observation is also comparable with that of the measurement by the dense surface detector array with a 750 m spacing of the Pierre Auger Observatory \cite{Abreu2021}. The spectrum will be successfully reconstructed up to $10^{19.5}\, {\rm eV}$ before the steep cutoff of the spectrum \cite{SOKOLSKY200967, PhysRevD.102.062005, COLEMAN2023102819}. Note that our prediction is made from a simple extrapolation of the best-fit power-law function to the TALE observation above $10^{17}\, {\rm eV}$. The measured CR energy spectrum has an ankle structure at around $10^{18.7}\, {\rm eV}$ where the spectrum becomes harder \cite{COLEMAN2023102819}. The expected number of CRs observed by GP300 in the ankle energy region should increase accordingly, and the ankle structure will be resolved. 

%\end{comment}

%\begin{table}[b!]
\begin{table}[h]
\centering
\begin{tabular}{cc}
\hline
${\rm log}_{10}(E[{\rm eV}])$ & Expected number of CR events per year\\
\hline
%$17.0-17.1$ & 0\\
%$17.1-17.2$ & 0\\
$17.2-17.3$ & 3821\\
$17.3-17.4$ & 8551\\
$17.4-17.5$ & 11457\\
$17.5-17.6$ & 8436\\
$17.6-17.7$ & 6034\\
$17.7-17.8$ & 3884\\
$17.8-17.9$ & 2402\\
$17.9-18.0$ & 1655\\
$18.0-18.1$ & 1055\\
$18.1-18.2$ & 624\\
$18.2-18.3$ & 378\\
$18.3-18.4$ & 244\\
$18.4-18.5$ & 142\\
$18.5-18.6$ & 91\\
$18.6-18.7$ & 56\\
$18.7-18.8$ & 33\\
$18.8-18.9$ & 21\\
$18.9-19.0$ & 12\\
$19.0-19.1$ & 7\\
$19.1-19.2$ & 4\\
$19.2-19.3$ & 2\\
$19.3-19.4$ & 1\\
$19.4-19.5$ & 1\\
\hline
\end{tabular}
\caption{Preliminary estimate of the number of CRs to be observed by GP300 in each logarithmic energy bin after one-year observation.}\label{tab1}
\end{table}

Finally, a very simple prediction of the statistical uncertainty in our future measurement of the mean depth of the air-shower maximum $X_{\rm max}$ is presented. LOFAR estimated the statistical uncertainty in their $X_{\rm max}$ measurement by $\sigma(X_{\rm max})/\sqrt{N}$, where $\sigma(X_{\rm max})$ is the standard deviation of the distribution of $X_{\rm max}$ in the energy bin of interest and $N$ is the number of events in that energy bin \cite{PhysRevD.103.102006}. Assuming similar event-reconstruction performance of GP300 to that of LOFAR and $\sigma(X_{\rm max}) = 60\, {\rm g}/{\rm cm}^2$ for the whole energy range accounting for the $X_{\rm max}$ measurement by LOFAR \cite{PhysRevD.103.102006}, the statistical uncertainty in mean $X_{\rm max}$ in the measurement by GP300 per energy bin ($\Delta {\rm ln}E = 0.1$) is estimated as $0.7$, $1.9$, and $6.5\, {\rm g}/{\rm cm}^2$ at $10^{17.5}$, $10^{18}$, and $10^{18.5}\, {\rm eV}$, respectively. The prediction of the statistical uncertainty is about five times smaller than those in the previous measurements using radio detectors, LOFAR and AERA \cite{PhysRevD.103.102006, PhysRevD.109.022002}, at $10^{17.5}\, {\rm eV}$. Therefore, systematic uncertainties will be a dominant contribution that limits the interpretation of the mean $X_{\rm max}$ measurement by GP300; for example the systematic uncertainty in the LOFAR measurement is $\simeq 7\, {\rm g}/{\rm cm}^2$ \cite{PhysRevD.103.102006}. The calibration of the DUs is a crucial work to reduce the systematic uncertainty.

\section{Conclusion}\label{sec3}
This study calculates the exposure of the GP300 radio detector array for CRs above $10^{17}\, {\rm eV}$. The trigger efficiency of GP300 is 50, 80, and 90\% at $10^{17.5}$, $10^{17.9}$, and $10^{18.3}\, {\rm eV}$, respectively. The trigger efficiency improves over the whole energy range above $10^{17}\, {\rm eV}$ with an increase of zenith angle $\theta$ in $65^{\circ} < \theta \lesssim 80^{\circ}$ due to the large footprint size of the radio emission on the ground. On the other hand, in the higher zenith-angle range in $\theta \gtrsim 80^{\circ}$, the trigger efficiency becomes smaller in $E \lesssim 10^{18}\, {\rm eV}$ due to the attenuation of the radio emission from very inclined showers.

The exposure of GP300 for one-day observation reaches $50$, $80$, and $90\, {\rm km}^2\, {\rm day}\, {\rm sr}$ at $10^{17.5}$, $10^{18}$, and $10^{19}\, {\rm eV}$, respectively. Based on the calculated exposure, about $130$ CR events are expected to be observed per day by GP300 in $E > 10^{17}\, {\rm eV}$ and $\theta > 65^{\circ}$. A total $\simeq 4.9\times 10^{4}$ CR events will be observed by GP300 after one-year observation, and the CR energy spectrum will be measured in $10^{17.2}\, {\rm eV} < E < 10^{19.5}\, {\rm eV}$ with a high statistical precision; the observation will have about eight and five times more statistics at around $10^{17.5}\, {\rm eV}$ and $10^{18}\, {\rm eV}$, respectively, compared with the previous measurement by the TALE fluorescence detector. Also, in the energy range of $10^{18}\, {\rm eV} \lesssim E \lesssim 10^{19}\, {\rm eV}$, the event statistics estimated for the GP300 observation is comparable with that of the measurement by the Pierre Auger Observatory using the dense surface detector array. Our future observation of mean $X_{\rm max}$ will also have a five times smaller statistical error than previous measurements using radio detectors at $10^{17.5}\, {\rm eV}$, and our interpretation of the chemical composition of CRs will be limited by systematic errors in our measurement. The observation of GP300 with a high statistical precision will surely help us accurately determine the spectral shape and the chemical composition of CRs in the transition region. 

%\begin{table}[b!]
%\centering
%\begin{tabular}{lccccccccccc}
%\hline
%Choice  &	Top1 & Top2 & Top3 & Top4 & Top5 & Top6 & Top7 & Top8 \\
% & Bottom1 & Bottom2 & Bottom3 & Bottom4 & Bottom5 & Bottom6 & Bottom7 & Bottom8 \\
%\hline
%X & 1 & 2 & 3 & 4 & 5 & 6 & 7 & 8 \\
%Y & a & b & c & d & e & f & g & h \\
%\hline
%\end{tabular}
%\caption{Table example.}\label{tab1}
%\end{table}

%\lipsum[3] 

% Bibtex references:
%\bibliographystyle{ICRC}
%\bibliography{references}

% Alternatively, you can include references by hand:
%\begin{thebibliography}{99}
%\bibitem{...}
%
%\end{thebibliography}

%\clearpage

%The following list of authors, affiliations and funding agencies will be updated at the day of submission. The following template is a placeholder generated with the member list updated to April 1, 2025..
\providecommand{\href}[2]{#2}\begingroup\raggedright\endgroup

\clearpage
\section*{Full Author List: GRAND Collaboration}

\scriptsize
\noindent
J.~Álvarez-Muñiz$^{1}$, R.~Alves Batista$^{2, 3}$, A.~Benoit-Lévy$^{4}$, T.~Bister$^{5, 6}$, M.~Bohacova$^{7}$, M.~Bustamante$^{8}$, W.~Carvalho$^{9}$, Y.~Chen$^{10, 11}$, L.~Cheng$^{12}$, S.~Chiche$^{13}$
% ...
\\
\\
$^{1}$Departamento de Física de Particulas \& Instituto Galego de Física de Altas Enerxías, Universidad de Santiago de Compostela, 15782 Santiago de Compostela, Spain \\
$^{2}$Institut d'Astrophysique de Paris, CNRS  UMR 7095, Sorbonne Université, 98 bis bd Arago 75014, Paris, France \\
$^{3}$Sorbonne Université, Université Paris Diderot, Sorbonne Paris Cité, CNRS, Laboratoire de Physique  Nucléaire et de Hautes Energies (LPNHE), 4 Place Jussieu, F-75252, Paris Cedex 5, France \\
$^{4}$Université Paris-Saclay, CEA, List,  F-91120 Palaiseau, France \\
$^{5}$Institute for Mathematics, Astrophysics and Particle Physics, Radboud Universiteit, Nijmegen, the Netherlands \\
$^{6}$Nikhef, National Institute for Subatomic Physics, Amsterdam, the Netherlands \\
$^{7}$Institute of Physics of the Czech Academy of Sciences, Na Slovance 1999/2, 182 00 Prague 8, Czechia \\
$^{8}$Niels Bohr International Academy, Niels Bohr Institute, University of Copenhagen, 2100 Copenhagen, Denmark \\
$^{9}$Faculty of Physics, University of Warsaw, Pasteura 5, 02-093 Warsaw, Poland \\
$^{10}$Key Laboratory of Dark Matter and Space Astronomy, Purple Mountain Observatory, Chinese Academy of Sciences, 210023 Nanjing, Jiangsu, China \\
$^{11}$School of Astronomy and Space Science, University of Science and Technology of China, 230026 Hefei Anhui, China \\
$^{12}$National Astronomical Observatories, Chinese Academy of Sciences, Beijing 100101, China \\
$^{13}$Inter-University Institute For High Energies (IIHE), Université libre de Bruxelles (ULB), Boulevard du Triomphe 2, 1050 Brussels, Belgium \\
% ...

%%%%%%%%%%%%%%%%%%%%%%%%%%%%%%%%%%%%%%%%%%%%%%%%%%%%%%%%%%%%%%
%%%%%%%%%%%%%%%%%%%%%%%%%%%%%%%%%%%%%%%%%%%%%%%%%%%%%%%%%%%%%%

\subsection*{Acknowledgments}

\noindent
The GRAND Collaboration is grateful to the local government of Dunhuag during site survey and deployment approval, to Tang Yu for his help on-site at the GRANDProto300 site, and to the Pierre Auger Collaboration, in particular, to the staff in Malarg\"ue, for the warm welcome and continuing support.
The GRAND Collaboration acknowledges the support from the following funding agencies and grants.
%%%%
\textbf{Brazil}: Conselho Nacional de Desenvolvimento Cienti\'ifico e Tecnol\'ogico (CNPq); Funda\c{c}ão de Amparo \`a Pesquisa do Estado de Rio de Janeiro (FAPERJ); Coordena\c{c}ão Aperfei\c{c}oamento de Pessoal de N\'ivel Superior (CAPES).
%%%%
\textbf{China}: National Natural Science Foundation (grant no.~12273114); NAOC, National SKA Program of China (grant no.~2020SKA0110200); Project for Young Scientists in Basic Research of Chinese Academy of Sciences (no.~YSBR-061); Program for Innovative Talents and Entrepreneurs in Jiangsu, and High-end Foreign Expert Introduction Program in China (no.~G2023061006L); China Scholarship Council (no.~202306010363); and special funding from Purple Mountain Observatory.
%%%%
\textbf{Denmark}: Villum Fonden (project no.~29388).
%%%%
\textbf{France}: ``Emergences'' Programme of Sorbonne Universit\'e; France-China Particle Physics Laboratory; Programme National des Hautes Energies of INSU; for IAP---Agence Nationale de la Recherche (``APACHE'' ANR-16-CE31-0001, ``NUTRIG'' ANR-21-CE31-0025, ANR-23-CPJ1-0103-01), CNRS Programme IEA Argentine (``ASTRONU'', 303475), CNRS Programme Blanc MITI (``GRAND'' 2023.1 268448), CNRS Programme AMORCE (``GRAND'' 258540); Fulbright-France Programme; IAP+LPNHE---Programme National des Hautes Energies of CNRS/INSU with INP and IN2P3, co-funded by CEA and CNES; IAP+LPNHE+KIT---NuTRIG project, Agence Nationale de la Recherche (ANR-21-CE31-0025); IAP+VUB: PHC TOURNESOL programme 48705Z. 
%%%%
\textbf{Germany}: NuTRIG project, Deutsche Forschungsgemeinschaft (DFG, Projektnummer 490843803); Helmholtz—OCPC Postdoc-Program.
%%%%
\textbf{Poland}: Polish National Agency for Academic Exchange within Polish Returns Program no.~PPN/PPO/2020/1/00024/U/00001,174; National Science Centre Poland for NCN OPUS grant no.~2022/45/B/ST2/0288.
%%%%
\textbf{USA}: U.S. National Science Foundation under Grant No.~2418730.
%%%
Computer simulations were performed using computing resources at the CCIN2P3 Computing Centre (Lyon/Villeurbanne, France), partnership between CNRS/IN2P3 and CEA/DSM/Irfu, and computing resources supported by the Chinese Academy of Sciences.

\end{document}